\documentstyle[epsf,rotate,preprint,aps,ifthen]{revtex}

\newboolean{intro}
\setboolean{intro}{true}

\begin{document}

\tighten

\title{Muon capture on Chlorine-35.}

\author{S. Arole $^1$, D.S. Armstrong $^2$, 
T.P. Gorringe $^1$, M.D. Hasinoff$^3$, 
M.A. Kovash$^1$, V. Kuzmin $^4$, B.A. Moftah $^{3 \dagger}$, 
R. Sedlar $^{1 \star}$, T.J. Stocki $^{3 \ddagger}$ 
and T. Tetereva $^5$.}

\vspace{4.0cm}

\address{$^1$ Department of Physics and Astronomy,
University of Kentucky,\\ Lexington, KY, USA 40506.} 

\address{$^2$ Department of Physics, College of William and Mary,
\\ Williamsburg, VA, USA 23187.}

\address{$^3$ Department of Physics and Astronomy, University of
British Columbia, \\ Vancouver, BC, Canada V6T 1Z1.}

\address{$^4$ Bogoliubov Laboratory of Theoretical Physics,
Joint Institute for Nuclear Research, Dubna, 141 980, Russia.}

\address{$^5$ Dubna Branch of Skobeltsyn Institute of
Nuclear Physics, Lomonosov Moscow State University,
Dubna, 141 980, Russia.}

\date{\today}

\maketitle


\abstract
{
We report measurements of 
$\gamma$--ray spectra from muon capture on $^{35}$Cl.
For the allowed Gamow--Teller transitions
to the $^{35}$S$( 2939 , 3/2^+ )$ state
and the $^{35}$S$( 3421 , 5/2^+ )$ state
we obtained their capture rates, hyperfine dependences
and $\gamma$--$\nu$ correlation coefficients.
The capture rates and hyperfine dependences were compared 
to shell model calculations 
using the complete 1s--0d space 
and the universal SD interaction.
With $g_p / g_a = 6.7$ and $g_a = -1.00$ (or $g_a = -1.26$)
we found agreement of the model and the data
at the 1--2 $\sigma$ level.
However, we caution that the transitions are sensitive
to $\ell = 2$ forbidden matrix elements.

\pacs{23.40.-s, 23.40.Hc, 27.30.+t}

\newpage

\section{Introduction}
\label{s-introduction}

The induced pseudoscalar coupling ($g_p$)
is the least known of 
the proton's weak couplings.
For the free proton, 
the coupling's determination 
is an important test 
of chiral symmetry breaking \cite{Go58,Be94,Fe97}.
For the bound proton,
the coupling's renormalization
is sensitive to $\pi$ exchange currents, $\Delta$--hole excitations 
and possible precursor effects of chiral phase transitions 
in hot, dense nuclear matter \cite{De76,De94,Rh84}.

Unfortunately the effects of $g_p$ are subtle and elusive.
In few-body systems recent results 
from radiative muon capture (RMC) on $^{1}$H \cite{Jo96a,Wr98}
and ordinary muon capture (OMC) on $^{3}$He \cite{Ac98}
are available.
However their interpretations are complicated
by muon chemistry in $^{1}$H
and 2--body currents in $^{3}$He,
and the puzzling discrepancy
between the values of $g_p$ from RMC on $^{1}$H
and OMC on $^{3}$He
is so far unexplained.
In complex nuclei new results 
from allowed transitions on 
$^{11}$B \cite{Wi98}, $^{23}$Na \cite{Jo96b}
and $^{28}$Si \cite{Br95,Mo97,Br00} are also available.
In nuclei the difficulty is disentangling the weak dynamics from
nuclear structure. 
The majority of data on nuclei are consistent with
an unrenormalized $g_p$,
but more experiments on other transitions
would be interesting.
 
In this article we report measurements
of $\gamma$--ray spectra from muon capture on $^{35}$Cl.
In particular we describe the determination of 
capture rates, hyperfine dependences and 
$\gamma$--$\nu$ angular correlations
for two $^{35}$Cl $\rightarrow$ $^{35}$S allowed 
Gamow--Teller transitions. 
We also compare their capture rates and hyperfine dependences
to a large basis shell model calculation,
and discuss their sensitivity to the weak couplings and the 
nuclear structure.

The paper is organized as follows:
In sections \ref{s-setup} and \ref{s-results}  
we describe the measurement setup and experimental results.
In section \ref{s-interpretation} we discuss the comparison
of the model and the data, 
and the sensitivity 
to the couplings constants 
and the nuclear model.
We conclude in section \ref{s-conclusion}.


\section{Experimental setup}
\label{s-setup}

The experiment was conducted 
using the M9B backward decay muon beamline 
at the TRIUMF cyclotron. 
The setup was similar to our earlier studies 
of other 1s--0d shell nuclei 
(see Refs. \cite{Jo96b,Mo97,Go99}).

We employed a $\mu^-$ beam of incident momentum
$60$~MeV/c, stopping rate $1.2 \times 10^5$~s$^{-1}$,
and $e$ and $\pi$ contamination of 10\% and $\leq$0.1\%.
Muon stops were counted 
in a plastic scintillator beam telescope comprising 
two counters ($S1$ and $S2$) upstream of the target
and one counter ($S3$) downstream of the target.
The target material was 
isotopically enriched Na$^{35}$Cl (99\% chlorine-35) powder
of mass $50$~g (we used isotopically pure chlorine-35
to avoid $\gamma$--ray backgrounds 
from $^{37}$Cl($\mu$,$2$$n$$\nu$) reactions). 
The material was packed in a thin--walled, disk--shaped,
polyethylene container.
The target was viewed by a HPGe detector
with an in-beam energy resolution of $2.2$~keV (full width half maximum)
and in-beam time resolution of $10$~ns (FWHM)
for $1.33$~MeV $^{60}$Co $\gamma$--rays.
A segmented NaI annulus, 
surrounding the HPGe detector,
was used to suppress the Compton scattering background.
A second NaI array,
viewing the Na$^{35}$Cl target,
was used for collection of $\gamma$--ray coincidence data.

Events were digitized on fulfillment of the logic condition
$Ge$$\cdot$$\overline{CS}$$\cdot$$\mu$$STOP$ $\cdot$$\overline{busy}$, where 
$Ge$ indicates a signal in the HPGe detector,
$\overline{CS}$ indicates no signal in the NaI suppressor,
$\mu$$STOP$ indicates a $\mu^-$ stop in the preceeding $2.0$~$\mu$s,
and $\overline{busy}$ indicates that the acquisition is live.
For each event we recorded energy and timing signals 
from the Ge detector, NaI arrays and beam counters.
We also recorded 
a multi--hit time history of muon stops,
and a pile--up bit and an overload bit from the HPGe.
The pile-up bit identifies events with a preceeding
Ge detector hit within 50~$\mu$s
and the overload bit identifies events with a preceeding 
$ > 10$~MeV Ge detector hit within 500~$\mu$s
(for details see Ref. \cite{Jo96b}).
Data were collected for
$4.2 \times 10^{10}$ muon stops in the Na$^{35}$Cl target, 
along with $\gamma$--ray background data from a LiCl target
and x--ray calibration data from various  materials.

Offline the data were sorted into histograms
corresponding to (i) a HPGe singles $\gamma$--ray energy spectra,
(ii) a HPGe$\cdot$NaI coincidence $\gamma$--ray energy spectra
({\it i.e.} a HPGe energy spectrum with a NaI coincidence requirement), 
and (iii) many HPGe$\cdot$$\mu$$STOP$ time--binned $\gamma$--ray 
energy spectra ({\it i.e.} HPGe energy spectra corresponding to
different time bins).
The time bins were defined by the time difference
between the incoming muon and the outgoing gamma-ray.
In filling the histograms we rejected any events 
with either an overload signal or a pile--up signal in the HPGe.

\section{Experimental results}
\label{s-results}

\subsection{HPGe calibration}
\label{s-calibration}

A large body of muonic x--ray data was collected in order to
determine the HPGe photon acceptance, energy resolution
and time resolution.
The data comprised K, L and M series x--ray spectra
from  S, Ca, In, Nb, Fe, Pb and Bi,
and spanned a range of energies from 500 to 4000~keV.

To determine the acceptance with muonic x-rays we
either (i) used published yield data for individual x--rays
\cite{Ke75,Ha76,Vo80,Ha82}
or (ii) assumed a yield of unity for the entire K--series.
Then we performed a least squares fit of the measured acceptances
to a smooth empirical curve to permit interpolation 
to the interesting $\gamma$--ray energies.

For the energy resolution determination we fit
the numerous x--ray peaks to a central Gaussian supplemented
by both low and high energy tails.
The width of the central Gaussian was observed to increase 
from 2.2~keV at 1000~keV to 3.6~keV at 3000~keV.
Both low energy and high energy tails were observed,
but their effects on the fits
to the $^{35}$S $\gamma$--rays 
were almost negligible 
(see Sec. \ref{s-correlations}).
The best fit values of the lineshape parameters 
were used to fix the instrumental lineshape 
in the least squares fits to the $\gamma$--ray data.

For the time resolution determination
we fit the prompt peaks of the x--ray lines 
with a central Gaussian supplemented
by both early and late tails 
(energy windows were used to select the x--rays).
We found that, for photon energies above 1000~keV,
the Gaussian width was approximately constant at 10~ns FWHM,
but the resolution rapidly got worse for lower energies. 
A weak tail at late times was observed, 
but its effect on the fits
to the $^{35}$S $\gamma$--rays 
was almost negligible 
(see Sec. \ref{s-hyperfine}). 
Again, the best fit values of the lineshape parameters
were used to fix the lineshape 
in the least squares fits to the $\gamma$--ray data.

More details on the procedures for the determination
of the photon acceptance, energy resolution and time resolution
can be found in the Refs. \cite{Jo96b,Go99}.

\subsection{Line Identification}
\label{s-identification}

About eighty peaks were found 
in the $\gamma$--ray spectrum from muon stops in Na$^{35}$Cl.
To match the peaks to known $\gamma$--ray lines we searched 
the nuclear $\gamma$--ray database
at the National Nuclear Data Center (NNDC) \cite{Ki96}
for all isotopes corresponding to the capture reactions
$^{23}$Na$( \mu^- , x $p~$ y $n~$ \nu)$ 
and $^{35}$Cl$( \mu^- , x $p~$ y $n~$ \nu)$ 
where $x, y  = 0, 1, 2$.
For conclusive identification we demanded 
(i) an energy match within experimental uncertainties,
(ii) consistency with other $\gamma$--branches of the parent state,
and (iii) Doppler broadening if the parent state lifetime
was less than the recoil stopping time.
This procedure yielded   
5 $\gamma$--rays from levels in $^{35}$S, 
21 $\gamma$--rays from levels in $^{34}$S,
and 4 $\gamma$--rays from levels in $^{33}$S. 

The conclusively identified $\gamma$--rays from $^{35}$S
are listed in Table \ref{t-gammas} and shown in Fig. \ref{f-gammas}. 
The five $\gamma$--rays all feed the $^{35}$S ground state
and correspond to three allowed transitions
to the $( 2717 , 5/2^+ )$, $( 2939 , 3/2^+ )$ and $(3421, 5/2^+)$ states 
and two forbidden transitions 
to the $( 1991 , 7/2^- )$ and $( 2439 , 3/2^- )$ states.

In addition, the production of $\gamma$--rays 
from the $^{35}$S$( 1572 , 1/2^+ )$ and 
$^{35}$S$( 4028 , 1/2^+ $--$ 5/2^+ )$ levels 
is possible.
For the $( 1572 , 1/2^+ )$ level, 
a $\gamma$--peak was clearly identified at $1572$~keV, 
but unfortunately it matches the energies of 
both the $1572 \rightarrow 0$ $^{35}$S transition 
and a $4877 \rightarrow 3304$ $^{34}$S transition. 
The observed $1572$~keV peak is Doppler broadened,
indicating that the short lifetime $^{34}$S state 
is the major $\gamma$--ray source.
However some contribution from the $^{35}$S$( 1572 , 1/2^+ )$ level 
is a possibility.
For the $( 4028 , 1/2^+ $--$ 5/2^+ )$ level, 
weak lines were identified at energies of $\sim$$5$~keV below
that expected for $4028 \rightarrow 1572$, $4028 \rightarrow 2348$
and $4028 \rightarrow 2939$ transitions.
Note the experimental uncertainty 
for the $4028$~keV excitation energy 
is about $\pm 2$~keV \cite{Ki96},
so these peaks in the Na$^{35}$Cl data 
may be evidence that the energy of the state
is actually about $4023$~keV.

\subsection{Line intensities}
\label{s-gamma yields}

To determine the gamma-ray yields per $\mu$Cl atom formed,
we employed the equation
\begin{equation}
{ Y_{\gamma} = {{\rm N_{\gamma}} 
\over {\rm N_{\mu}} ~~\epsilon \Delta \Omega (E) ~~f_{iso}}
~~C_{ab} ~~C_{sv}  ,}
\label{e-gamma yields}
\end{equation}
where N$_{\gamma}$ is the number of $\gamma$--rays detected,
N$_{\mu}$ is the number of livetime-corrected muon stops,
$\epsilon \Delta \Omega (E)$ is the photon acceptance 
at the appropriate energy,
$f_{iso}$ is the $\mu^-$ Cl atomic capture fraction,
and $C_{ab}$ and $C_{sv}$ are minor correction factors,
described below.
The $\gamma$--ray counts N$_{\gamma}$
were obtained from fits with Doppler broadened or Gaussian lineshapes
(see section \ref{s-calibration}).
The atomic capture fraction $f_{iso} = 0.59 \pm 0.04$, {\it i.e.} 
the fraction of muon stops in NaCl that undergo atomic capture on Cl,
was taken from Ref.\ \cite{Ma77}\footnote{For 
comparison the Fermi Z-law yields $f_{iso} = 0.61$
and the calculation of Vogel {\it et al.}\ \cite{Vo75} yields  
$f_{iso} = 0.58$. The earlier experimental work
of Knight {\it et al.}\ \cite{Kn76} gave $f_{iso} = 0.56$
and Zinov {\it et al.}\ \cite{Zi66} gave $f_{iso} = 0.49$.}.
The factors $C_{ab}$ and $C_{sv}$ 
accounted for photon absorption in the target 
and self-vetoing by the suppressor
($C_{ab}$ varied from $1.02$ to $1.03$
and $C_{sv}$ varied from $1.06$ to $1.16$).
They are discussed in detail in Ref.\ \cite{Jo96b}.

The resulting gamma-ray yields
per $\mu$Cl atom formed
are given in Table \ref{t-gammas}. 
The dominant uncertainties
were the measurement uncertainty in the atomic capture fraction 
of $\pm$7\%
(see Ref.\ \cite{Ma77} for details)
and the normalization uncertainty in the muonic x-ray calibration 
of $\pm$15\% 
(see Ref.\ \cite{Jo96b} for details).
The statistical uncertainties in N$_{\gamma}$
and total uncertainties in $C_{ab}$ and $C_{sv}$ 
were negligible.

\subsection{Cascade feeding}
\label{s-feeding}

A serious concern in $\gamma$--ray studies is cascade feeding into 
interesting $^{35}$S levels from higher--lying levels.
If missed, such production of $^{35}$S states
would distort the interpretation 
of the measured rates, hyperfine dependences and
angular correlations.

Although the comparison between 
our Na$^{35}$Cl data and the NNDC Tables \cite{Ki96}
gave only five clean $\gamma$--ray matches,
some $\gamma$--peaks were never identified
and many $^{35}$S states have unknown decays.
Therefore we performed a $\gamma$--$\gamma$ coincidence measurement
in order to study the total amount of cascade feeding to $^{35}$S levels.
Specifically,
we determined the counts for  
interesting $\gamma$--ray lines 
in the HPGe singles spectrum (denoted N$_{S}^{\gamma}$)
and the HPGe$\cdot$NaI coincidence spectrum (denoted N$_{C}^{\gamma}$),
and computed the super-ratio
\begin{equation}
\label{e:f}
f_c = {N_{C}^{\gamma} / N_{S}^{\gamma}
\over N_{C}^{Co60} / N_{S}^{Co60}}
\end{equation}
The coicidence fraction $f_c$ reflects the total number 
of other $\gamma$--rays in prompt coincidence 
with the interesting $\gamma$--ray.
Note in Eqn.\ \ref{e:f} the ratio $N_{C}^{Co60} / N_{S}^{Co60}$ 
was obtained by measuring the well known 1.17 and 1.33 MeV 
coincident $\gamma$--rays from a Co-60 source.
The ratio  $N_{C}^{Co60} / N_{S}^{Co60}$
serves to normalize the ratio $N_{C}^{\gamma} / N_{S}^{\gamma}$.
We stress that the approach has limitations:
(i) it cannot distinguish feeding 
via a single--step cascade 
from a multi--step cascade,
and (ii) its interpretation assumes an energy-independent acceptance 
for the $\gamma$--ray detection by the NaI array 
(the latter assumption is fairly reasonable for $\gamma$--ray energies
from 0.5 to 3.0~MeV).
The method is described in detail in Refs.\ \cite{Jo96b,Mo97}.

The measured coincidence fractions $f_c$ for
$^{35}$S $\gamma$--rays are listed in column seven of Table \ref{t-gammas}.
For the 1991, 2348 and 2717~keV $\gamma$--rays
the table shows the presence of considerable feeding 
from unidentified levels.
The comparison of experiment and theory 
is therefore not warranted in such cases.
However for the 2939 and 3421~keV $\gamma$--rays
the table shows the absence of large amounts of cascade feeding.
Therefore we focused our attention
on the model-data comparison of the physical observables
in the $^{35}$Cl$(3/2^+ , 0)$ $\rightarrow$ 
$^{35}$S$( 3/2^+ , 2939)$ transition
and the $^{35}$Cl$(3/2^+ , 0)$ $\rightarrow$ 
$^{35}$S$( 5/2^+ , 3421)$ transition.

\subsection{Capture rates}
\label{s-rates}

To convert the gamma yields to capture rates
we accounted for the 
$\gamma$--ray branching ratios (taken from Ref. \cite{Ki96})
and multiplied by the 
muon disappearance rate (taken from Ref. \cite{Su87}).
For $^{35}$Cl$(3/2^+ , 0)$ $\rightarrow$ $^{35}$S$( 3/2^+ , 2939)$
and $^{35}$Cl$(3/2^+ , 0)$ $\rightarrow$ $^{35}$S$( 5/2^+ , 3421)$
the resulting rates and experimental uncertainties
are listed in Table \ref{t-results}.

Because of the hyperfine effect in the $\mu^-$Cl atom,
the experimental capture rates are combinations 
of hyperfine capture rates 
(for details see Sec. \ref{s-hyperfine}).
Assuming an initial statistical population 
of the two hyperfine states,
the hyperfine transition rate of Sec. \ref{s-hyperfine},
the published muon disappearance rate of Ref. \cite{Su87},
and a $2.0 \mu$s wide $\mu^-$STOP gate, 
the observed rates $\Lambda$ in Table \ref{t-results} are 
related to the hyperfine capture rates via 
$\Lambda = 0.12 \Lambda_+ + 0.88 \Lambda_-$.
The extraction of the hyperfine capture ratio 
$\Lambda_+ / \Lambda_-$ is discussed in
the following section.

\subsection{Hyperfine dependences}
\label{s-hyperfine}

In a non--zero spin $I \neq 0$ target the muonic atom's
1S atomic state is split into hyperfine states 
with F$_+$~$=$~$I + 1/2$ and F$_-$~$=$~$I - 1/2$.
Due to the spin dependence of the weak interaction
the capture rates from the hyperfine states are usually different.
In certain cases ({\it e.g.} $^{35}$Cl),
during the muonic atom lifetime 
the upper HF state decays to the lower HF state. 
Consequently, the time dependence 
of $\gamma$--rays from $\mu^-$ capture has the form
\begin{equation}
 A e^{- \Lambda_D t}(1 + k e^{- \Lambda_h t}) ,
\label{e-time}
\end{equation}
where $\Lambda_D$ is the $\mu^-$ disappearance rate,
$\Lambda_h$ is the hyperfine transition rate, 
and $k$ is related to the hyperfine dependence
of muon capture via $k = f_+ (\Lambda_+$/$\Lambda_- - 1)$.
The quantity $f_+$ 
is the fractional population of the $F_+$  state at $t = 0$. 
For a spin $3/2$ target and a statistical HF population \cite{Co93}
the factor is $f_+ = 5/8$.
Consequently, if $\Lambda_h$ 
is neither too fast nor too slow
({\it i.e.} $\Lambda_h$ $\sim$ $\Lambda_D$),
the hyperfine dependence of muon capture 
can be determined from the time spectrum of the $\gamma$--rays.

For each $\gamma$-ray line its time-binned energy spectra
were first fit to appropriate Gaussian or Doppler lineshapes
in order to determine its time spectrum.
The $\gamma$--ray time spectra themselves
were then fit to a convolution of the 
theoretical time dependence (Eqn. \ref{e-time})
with the measured HPGe time resolution (Sec. \ref{s-calibration}).
We accounted for the slight distortion of the disappearance rate
due to muon pile--up in the Na$^{35}$Cl target
by the method described in Ref. \cite{Jo96b} (a 4\% effect).
Additionally we investigated the sensitivities
of $\Lambda_+$/$\Lambda_-$ and $\Lambda_h$ 
to the various parameters 
of the instrumental response and the muon pile-up.

In this analysis 
we employed the 1991~keV $\gamma$--ray time spectrum
to determine the hyperfine transition rate $\Lambda_h$
(since the 1991~keV $\gamma$--ray has a very large hyperfine effect).
Note,
our best fit value of $\Lambda_h = 10.1 \pm 1.0$~$\mu s^{-1}$ 
for this NaCl experiment 
and the earlier value of $\Lambda_h = 8.1 \pm 2.2$~$\mu s^{-1}$
for a LiCl experiment \cite{Go93},
are consistent
(we are also consistent with the recent experiment
of Stocki {\it et al.}\ \cite{St01}).
Also our experimental rate is in reasonable agreement with the predicted rate
of $\Lambda_h = 8.0$~$\mu s^{-1}$ from Ref. \cite{Wi63}.

The best fit values for the hyperfine dependences of the
2939~keV line and 3421~keV line are given in Table \ref{t-results}
(in extracting $\Lambda_+ / \Lambda_-$ we fixed $\Lambda_h$
using the 1991~keV $\gamma$--ray).
In addition, 
representative fits to their $\gamma$--ray time spectra
are shown in Fig. \ref{f-timefits}
and representative sensitivities to the various fitting parameters 
are shown in Tables \ref{t-2939time} and \ref{t-3421time}.
The quoted errors for $\Lambda_+ / \Lambda_-$
include the statistical errors in the fitting procedure
and the sensitivities to the determination of the
instrumental resolution and the muon pile--up.

\subsection{Angular correlations}
\label{s-correlations}

Because of the spin dependence of the weak interaction,
the recoil produced in muon capture is usually oriented.
In general, this recoil orientation yields an angular correlation between the 
neutrino direction and the decay $\gamma$--ray direction.
Further,
when the $\gamma$--decay lifetime is less than the recoil stopping time,
the $\gamma$--$\nu$ directional correlation will be manifest in the 
$\gamma$--ray Doppler spectrum. 
Both the 2939~keV $\gamma$--ray and the 3421~keV $\gamma$--ray
were Doppler broadened.

For unpolarized muons the directional correlation 
of the photon and the neutrino 
has the form
\begin{equation}
1 + \alpha P_2 ( \cos{ \theta } ) , 
\end{equation}
where $\theta$ is the angle between the momentum vectors
of the photon and the neutrino,
$P_2 ( \cos{ \theta } )$ is the Legendre polynomial,
and $\alpha$ is the $\gamma$--$\nu$ angular correlation coefficient. 
Note, for recoil spins  $J \leq 2$
or $\gamma$--ray multipolarities $L \leq 2$,
higher-order Legendre polynomials do not contribute.

To fit the $\gamma$--ray Doppler spectra,
and extract the $\gamma$--$\nu$ correlation coefficient,
we convoluted the theoretical Doppler spectrum
with the HPGe instrumental lineshape. 
In the fitting procedure 
we fixed the lineshape parameters at their known values,
and studied the background sensitivity by trying background shapes
of various forms ({\it e.g.} linear, quadratic and exponential).  
Representative fits to the 2939~keV and 3421~keV lines
are shown in Figs. \ref{f-dop2939} and \ref{f-dop3421}
and typical sensitivities to the fitting parameters
are listed in Tables \ref{t-dop2939} and \ref{t-dop3421}.
For the $2939$~keV line a concern is clearly the
background line at 2935~keV,
which obscures a portion of the Doppler broadened spectrum 
of the $2939$~keV gamma-ray. 

Unfortunately the Doppler lineshape 
may be distorted
by the slowing down
of the recoil ion.
Specifically, if the slowing-down time $t_s$ and the gamma-ray lifetime $\tau$
are comparable, then the lineshape is a function
of the coefficient $\alpha$ and the ratio $\tau / t_s$.
From sample fits with different lifetimes
we found the input value of $\tau / t_s$
and output value of $\alpha$ were highly correlated
when $\tau / t_s > 0.05$ (see Figs. \ref{f-alpha2939} and \ref{f-alpha3421}).
For $\tau / t_s < 0.01$ the $2939$~keV line yielded $\alpha = -0.43 \pm 0.13$
and the $2939$~keV line yielded $\alpha = -0.39 \pm 0.05$
(the quoted errors for the coefficients 
include the statistical uncertainties from the fitting procedure
and the various sensitivities to the fitting parameters).
However, for $\tau / t_s > 0.01$ the magnitude of $\alpha$ decreases
as the value of $\tau / t_s$ increases. 
Therefore,
since independent determination  
of the gamma-ray lifetimes 
are currently unavailable,
our results for $\alpha$ are functions of $\tau / t_s$ 
({\it i.e.} the plots of Fig. \ref{f-alpha2939} and Fig. \ref{f-alpha3421}).
When the necessary lifetimes are finally measured 
the correlation coefficients
may be read off these figures.

Finally because of the hyperfine effect,
the observed coefficient $\alpha$
is a linear combinations 
of the two coefficients $\alpha_{\pm}$
of the two $F_{\pm}$ states.
Specifically the observed correlation $\alpha$ is
(see Ref. \cite{Ci84} for details)
\begin{equation}
\alpha = {n_+ \Lambda_+  \alpha_+ + n_- \Lambda_- \alpha_- \over
n_+ \Lambda_+  + n_- \Lambda_- } ,
\end{equation}
where $n_+/n_-$
is the relative muon occupancy of the hyperfine states
and $\Lambda_+$/$\Lambda_-$
is the relative capture rates from the hyperfine states.
The relative occupancy was taken from Sec. \ref{s-hyperfine},
and is
$n_+/n_- = 0.12/0.88 = 0.14$.
The hyperfine dependences were taken from Table \ref{t-results},
and are
$0.74 \pm 0.17$ for the $( 2939 , 3/2^+ )$ transition
and $1.60 \pm 0.19$ for the $( 3421 , 5/2^+ )$ transition.
This yields $\alpha = 0.09 \alpha_+ + 0.91 \alpha_-$ 
for the $( 2939 , 3/2^+ )$ transition and
$\alpha = 0.18 \alpha_+ + 0.82 \alpha_-$ 
for the $( 3421 , 5/2^+ )$ transition.

\section{Interpretation}
\label{s-interpretation}

Herein we discuss the comparison of experiment and theory
for the measured observables
in the allowed transitions 
to the $( 2939 , 3/2^+ )$ level 
and the $( 3421 , 5/2^+ )$ level.
In Sec. \ref{s-general} we describe the general dependences 
on weak dynamics,
and in Sec. \ref{s-model} we discuss 
the model calculation and input parameters.
Sec. \ref{s-comparisons} compares the
model results with the experimental data,
and emphasizes the sensitivities 
to the weak couplings
and the nuclear structure.

Note that we restrict the comparison of theory and experiment
to the capture rates and the hyperfine dependences.
Interpretation of the correlations
is not possible at the moment
as the $\gamma$-decay lifetimes
and their mixing ratios
are not currently available.

\subsection{Fujii-Primakoff Approximation}
\label{s-general}


To demonstrate the basic features of the physical observables
in allowed $3/2^+ \rightarrow 3/2^+$ and $3/2^+ \rightarrow 5/2^+$ transitions
we first review some results 
of the Fujii--Primakoff approximation \cite{Fu59}.
Recall that the Fujii--Primakoff Hamiltonian 
for nuclear muon capture \cite{Pr59}
is
\begin{equation}
\label{e: FP hamiltonian}
H = \tau^+ {\small {1 - \sigma \cdot \hat{\nu} \over 2}} 
\sum_{i = 1}^{A} \tau^-_i
( 
G_V~ 1 \cdot 1_i
+ G_A~ \sigma \cdot \sigma_i 
+ G_P~ \sigma \cdot \hat{\nu} ~ \sigma_i \cdot \hat{\nu}
)
\, \delta ( r - r_i ) ,
\end{equation}
where $\hat{\nu}$ 
is the $\nu$--momentum unit vector,
$1$ and $\sigma$ are unit and spin matrices
(the operators with subscripts act on nucleons
and the operators without subscripts act on leptons),
and $\tau^+$ converts the muon into a neutrino.
Note that the leading contribution of $g_a$ is to the effective coupling $G_A$
and the leading contribution of $g_p$ is to the effective coupling $G_P$.
In the Fujii-Primakoff approximation 
only the allowed GT operator
and the allowed Fermi operator
are retained following the multipole expansion
of Eqn. \ref{e: FP hamiltonian}.


First consider the hyperfine dependence of muon capture.
We note in the multipole expansion of the Fujii-Primakoff Hamiltonian
the $G_A$--term 
makes allowed contributions to neutrino waves 
with total angular momentum $j^{\pi} = 1/2^+$ only
whereas the $G_P$--term 
makes allowed contributions to neutrino waves
with total angular momentum $j^{\pi} = 3/2^+$
(this difference is because 
of $\hat{\nu}$ in Eqn. \ref{e: FP hamiltonian}).
In $3/2^+ \rightarrow 5/2^+$  transitions, 
neutrinos with $j = 1/2^+$ may be emitted in $F_+$ capture,
but neutrinos  with $j = 3/2^+$ must be emitted in $F_-$ capture.
This makes $\Lambda_+ / \Lambda_- >> 1$ 
and a strong function of the ratio $g_p / g_a$.
However in $3/2^+ \rightarrow 3/2^+$  transitions 
the emission of $j = 1/2^+$ neutrinos is possible
for both $F_+$ atoms and $F_-$ atoms.
This makes $\Lambda_+ / \Lambda_- \sim 1$ 
and a weaker function of the ratio $g_p / g_a$.


Next consider the $\gamma$--$\nu$ correlation in muon capture.
Note in Eqn. \ref{e: FP hamiltonian} 
the $G_A$--term is a vector in spin--space
while the $G_P$--term is a scalar in spin--space.
Consequently, in most cases the magnetic sub-states of recoil nuclei
are populated differently by allowed contributions 
that originate from the $G_A$-term and the $G_P$-term.
This makes the recoil orientation, 
and therefore the $\gamma$-$\nu$ correlation,
a function of $g_p / g_a$.
However an exception is $\gamma$--$\nu$ correlations
in $3/2^+ \rightarrow 5/2^+$ transitions
on $F_-$ atoms. 
In such circumstances the emission of d--wave $\nu$'s is required
and only the $G_P$--term generates an allowed contribution.
Consequently 
for $3/2^+ \rightarrow 5/2^+$ transitions
on $F_-$ atoms 
the recoil orientation and $\gamma$-$\nu$ correlation
is independent of the coupling constants in the Fujii-Primakoff
approximation.
The potential usefulness of the $\gamma$-$\nu$ correlation coefficients
for model testing is illustrated further in Appendix \ref{s-appendix}.

In summary, different observables offer different sensitivities
to the weak couplings and the nuclear structure.
For example, the hyperfine dependence $\Lambda_+ / \Lambda_-$
in a $3/2^+ \rightarrow 5/2^-$ transition is particularly sensitive
to the induced pseudoscalar coupling
and the correlation coefficient $\alpha_-$
in a $3/2^+ \rightarrow 5/2^-$ transition is particularly sensitive
to the forbidden matrix elements.
In principle they enable a means of
both extracting the weak couplings
and testing the nuclear structure.

\subsection{Model calculation}
\label{s-model}


Our model calculations of the 
capture rates and hyperfine dependences
were performed in the impulse approximation using the shell model
(equations for the capture rates and the hyperfine dependences 
are published in Refs. \cite{Wa75,Mo60,Ci84}).
Specifically, 
we used the computer code OXBASH \cite{Ox86}, 
the complete 1s--0d space,
and the universal SD interaction \cite{Wi84}.
We fixed the weak vector and magnetic couplings 
to the values  $g_ v = 1.000$ and $g_m = 3.706$ 
and varied the weak axial and induced pseudoscalar couplings.
The $A = 35$ nuclear matrix elements 
were computed with harmonic oscillator wavefunctions
and an oscillator parameter $b = 1.90$~fm.
The momentum transfer was computed using
the experimental values of excitation energies
(giving $q = 101.17$~MeV/c for the 2939~keV state
and $q = 100.59$~MeV/c for the 3421~keV state).
Finally the $\mu^-$$^{35}$Cl atomic wavefunction 
was assumed to be uniform in the nuclear volume 
and computed employing a muon wavefunction reduction factor of $R =0.521$
(see Walecka \cite{Wa75} and references therein).
More details are given in Ref. \cite{Jo96b}.


The arguments we made in Sec. \ref{s-general} 
for sensitivities of observables 
to $g_a$ and $g_p$  
were based
on the dominance
of the allowed GT matrix element 
in the interesting $( \mu , \nu )$ transition.
However our model calculations 
show a large
$(0d_{5/2})^{12}(1s_{1/2})^{4}(0d_{3/2})^3$ component
in the $^{35}$Cl ground state
and a large
$(0d_{5/2})^{12}(1s_{1/2})^{3}(0d_{3/2})^4$ component
in the two $^{35}$S states.
Consequently the 
$1s_{1/2} \rightarrow 0d_{3/2}$ single particle transition
is important
in both $^{35}$Cl$(3/2^+ , 0)$ $\rightarrow$ $^{35}$S$( 3/2^+ , 2939)$
and $^{35}$Cl$(3/2^+ , 0)$ $\rightarrow$ $^{35}$S$( 5/2^+ , 3421)$.
Furthermore for $1s_{1/2} \rightarrow 0d_{3/2}$ single particle transitions
the allowed GT matrix element is zero
(the operator obeying a $\Delta \ell = 0$ selection rule).
This amplifies the importance of 
$\ell = 2$ forbidden contributions
in both transitions.
A serious concern is clearly therefore the 
correct accounting
for $\ell = 2$ forbidden contributions
from the $1s_{1/2} \rightarrow 0d_{3/2}$ single particle transition
in the $^{35}$Cl$(3/2^+ , 0)$ $\rightarrow$ $^{35}$S$( 3/2^+ , 2939)$
and the $^{35}$Cl$(3/2^+ , 0)$ $\rightarrow$ $^{35}$S$( 5/2^+ , 3421)$
\footnote{The full calculation 
shows numerous other configurations
with additional holes 
in the $0d_{5/2}$--$1s_{3/2}$ orbitals
of the relevant $A = 35$ states. 
These generate single particle contributions
in $^{35}$Cl$(3/2^+ , 0)$ $\rightarrow$ $^{35}$S$( 3/2^+ , 2939)$
and $^{35}$Cl$(3/2^+ , 0)$ $\rightarrow$ $^{35}$S$( 5/2^+ , 3421)$
involving $0d \rightarrow 0d$
and $1s \rightarrow 1s$ transitions.
Such single particle transitions
generate substantial allowed Gamow-Teller contributions.}.


We further note that the calculation  
shows that the $^{35}$Cl $\rightarrow$ $^{35}$S GT strength
is spread over many states with $E_x < 10$~MeV
(see Fig. \ref{f-bgt}).
This contrasts with $^{23}$Na $\rightarrow$ $^{23}$Ne
and $^{28}$Si $\rightarrow$ $^{28}$Al
(other cases of $\mu$ capture work on 1s--0d nuclei)
where only a few states are found to exhaust a large fraction
of GT strength (for more details 
see Gorringe {\it et al.}\ \cite{Go99}).

\subsection{Comparison of experiment and theory}
\label{s-comparisons}


Fig. \ref{f-rate} shows the results of our calculations
of the capture rate $\Lambda = 0.12 \Lambda_+ + 0.88 \Lambda_-$ 
to the $( 2939 , 3/2^+ )$ state
and the $( 3421 , 5/2^+ )$ state.
As expected we found
that the rate is most sensitive 
to the value of the coupling $g_a$.
For $g_a = -1.26$ and $g_p / g_a = 6.8$
the computed rates 
were $12.2 \times 10^3~$s$^{-1}$ for the $( 2939 , 3/2^+ )$ state
and $15.4 \times 10^3$~s$^{-1}$ for the $( 3421 , 5/2^+ )$ state.
For $g_a = -1.00$  and $g_p / g_a = 6.8$
the computed rates 
were $8.4 \times 10^3$~s$^{-1}$ for the $( 2939 , 3/2^+ )$ state
and $12.3 \times 10^3$~s$^{-1}$ for the $( 3421 , 5/2^+ )$ state.
The calculated rates are remarkably close 
to the corresponding measurements
of $( 12.2 \pm 2.2 ) \times 10^3$~s$^{-1}$ 
and $( 11.9 \pm 2.2 ) \times 10^3$~s$^{-1}$
respectively.

Concerning the induced pseudoscalar coupling, 
the calculated $( 3421 , 5/2^+ )$ rate 
was found to exhibit some sensitivity to $g_p$
but the calculated $( 2939 , 3/2^+ )$ rate 
was found to exhibit no sensitivity to $g_p$
(see Fig. \ref{f-rate}).
This finding is in agreement with the expectations
of the Fujii-Primakoff Approximation
(see Sec. \ref{s-general}).


Figs. \ref{f-hyp} show the results of our calculations
for the hyperfine dependence $\Lambda_{+} / \Lambda_{-}$ 
of the $( 2939 , 3/2^+ )$ transition
and the $( 3421 , 5/2^+ )$ transition.
Note as argued in Sec. \ref{s-general},
the $3/2^+ \rightarrow 5/2^+$ hyperfine effect
is quite strongly dependent on $g_p$
whereas the $3/2^+ \rightarrow 3/2^+$ hyperfine effect
is quite weakly dependent on $g_p$.
For $g_p / g_a = 6.7$
the calculation 
gives $\Lambda_{+} / \Lambda_{-} = 1.3$--$1.4$
for the $( 3421 , 5/2^+ )$ transition
and $\Lambda_{+} / \Lambda_{-} = 0.45$--$0.46$
for the $( 2939 , 3/2^+ )$ transition
(the range corresponds to choosing either $g_a = -1.00$ or $g_a =-1.26$).
For comparison the experimental values  
are $1.60 \pm 0.19$ and $0.74 \pm 0.17$
respectively.

Clearly the model calculations and experimental results for
hyperfine dependences are in semi-quantitative agreement;
for example both concurring
that $F_-$ capture is stronger for the $( 2939 , 3/2^+ )$ transition
and $F_+$ capture is stronger for the $( 3421 , 5/2^+ )$ transition.
However,
at the level of 1--2 $\sigma$, 
some indication of disagreement between model and data
is suggested for $g_p / g_a = 6.7$,
the data favoring a smaller value of $g_p / g_a$.

Is this evidence for the medium modification
of the induced coupling?
Recall in discussing the $A =35$ structure 
in Sec. \ref{s-model} we identified
the large contribution 
of $1s_{1/2} \rightarrow 0d_{3/2}$ transitions
and $\ell = 2$ forbidden operators.
For example,
the fact that $\Lambda_+ \sim \Lambda_-$ 
not $\Lambda_+ >> \Lambda_-$
for $^{35}$Cl($3/2^+$ , 0) $\rightarrow$ $^{35}$S( $5/2^+$ , 3421),
both in the data and in the model,
is supporting evidence 
that forbidden terms
are important here.
This means that the observables are sensitive to the ratio
between the allowed Gamow--Teller matrix elements
and the various $\ell = 2$ forbidden matrix elements.
The matrix elements are  rather dependent 
on the particular admixtures
of the various hole--states
in the leading $(0d_{5/2})^{12}(1s_{1/2})^{4}(0d_{3/2})^3$ configuration 
for $^{35}$Cl
and the leading $(0d_{5/2})^{12}(1s_{1/2})^{4}(0d_{3/2})^3$ configuration 
for $^{35}$S.
Therefore we believe it is unwise to blame 
the coupling $g_p$ for the small discrepancies 
between the model and the data.

We should also remind the reader that some cascade feeding
into the $( 2939 , 3/2^+ )$ state 
and the $( 3421 , 5/2^+ )$ state
is possible (see Sec. \ref{s-feeding}).
If feeding is present, 
the measured capture rate
will over--estimate the direct capture rate,
and the measured hyperfine dependence 
may differ from the direct hyperfine dependence.
This also could account 
for some remaining discrepancies between experiment and theory.

\section{Conclusion}
\label{s-conclusion}

In summary, we have measured the $\gamma$--ray spectra 
from muon capture on isotopically enriched Na$^{35}$Cl.
For 
the $^{35}$Cl($3/2^+$ , 0) $\rightarrow$ $^{35}$S( $3/2^+$ , 2939) transition
and $^{35}$Cl($3/2^+$ , 0) $\rightarrow$ $^{35}$S( $5/2^+$ , 3421) transition
we obtained their capture rates, hyperfine dependences
and $\gamma$-$\nu$ correlation coefficients.
Concerning the capture rates and hyperfine dependences the 
experimental results 
are in semi--quantitative agreement
with a large basis shell model calculation
using $g_p / g_a = 6.7$
and $g_a = -1.00$ (or $g_a = -1.26$).
However we note
that large $\ell = 2$ forbidden contribution
from large $1s_{1/2} \rightarrow 0d_{3/2}$ single particle transitions
are probable indications of model dependences.
We therefore are unwilling 
to claim a determination of $g_p$ nor $g_a$,
with any precision.

For the angular correlations,
since independent measurements of gamma-ray lifetimes and mixing ratios
are currently unavailable,
the comparison of model and data was thwarted.
With such supplemental $\gamma$--ray data
the correlation coefficients for
the two $\gamma$-rays
would permit additional testing
of the calculation
and the extraction
of the coupling $g_p$.
For further details see Sec.\ \ref{s-correlations} 
and Appendix \ref{s-appendix}.
We therefore encourage any future efforts
to measure these quantities.

We would also encourage
a more thorough theoretical investigation 
of these allowed Gamow-Teller transitions.
In particular,
a more quantitative assessment
of the model uncertainties
and the various approximations
is worthwhile.
For example, 
in our model we have employed 
a uniform muon wavefunction
and harmonic oscillator nuclear wavefunctions,
and the effect of these simplifications
on the observables should be studied.


We wish to thank the staff 
of TRIUMF for their support.
In addition we thank both Prof. Jules Deutsch
and Prof. Harold Fearing for helpful discussions,
and the National Science Foundation (United States) 
and the Natural Sciences and Engineering Research Council (Canada) 
for financial assistance.

\appendix

\section{Usefulness of the $\gamma$-$\nu$ correlations.}
\label{s-appendix}

In this appendix we describe the usefulness 
of the $\gamma$-$\nu$ correlation coefficients 
in the testing of the nuclear model calculation.
We consider the example of the 3421~keV gamma-ray
from the $( 5/2+ , 3421)$ excited state.
Assuming negligible slowing-down effects
the measured $\gamma$-$\nu$ correlation coefficient 
was found to be
$\alpha = -0.39 \pm 0.05$.

In our toy model for the 
$^{35}$Cl$(3/2^+ , 0)$ $\rightarrow$ $^{35}$S$( 5/2^+ , 3421)$
$\rightarrow$ $^{35}$S$( 3/2^+ , 0)$ sequence we
will assume that the $^{35}$S( $5/2^+$ , 3421) lifetime
is fast enough that slowing-down effects are negligible
and that the 3421~keV gamma-ray is pure M1 radiation.
The purpose of the toy model is to illustrate the
potential sensitivities of the correlation $\alpha$,
and the importance of a measurement
of the $^{35}$S$( 5/2^+ , 3421)$ lifetime and
the 3421~keV E2/M1 mixing ratio. 

According to Ciechanowicz and Oziewicz \cite{Ci84}
the $\gamma$-$\nu$ angular correlation coefficient $\alpha$
may be written as the product 
\begin{equation}
\alpha = a_2 B_2
\end{equation}
where $a_2$ is determined by the $\mu$ capture process 
and $B_2$ is determined by the $\gamma$-decay process.
We obtained $a_2$ using the 1s-0d shell model
with the universal SD interaction as discussed 
in Sec.\ \ref{s-model}.
We took $B_2 = +0.3741 $ from Table 1 of 
Ciechanowicz and Oziewicz \cite{Ci84}
under the assumption
of a pure M1 decay.
Note that we computed the two correlations coefficients
$\alpha^{\pm}$ for the two hyperfine states F$_{\pm}$
and then combined the values to obtain
the observed correlation $\alpha = 0.18 \alpha_+ + 0.82 \alpha_-$
(see Sec.\ \ref{s-correlations} for details).

Our toy model results are given in 
Table \ref{t-toy} for $g_a = -1.26$.
It shows results from the full calculation 
as well as from a calculation omitting
the interesting second forbidden matrix elements.
As argued in Sec.\ \ref{s-general}
the correlation coefficient
is rather weakly dependent on the induced coupling $g_p$
but somewhat more dependent on second forbidden terms.

Our measured value $\alpha = -0.35 \pm 0.05$ 
and the calculated values $\alpha \simeq -0.23$
are significantly different.
Unfortunately in the absence of any experimental
data on the state lifetime and the E2/M1 mixing
ratio it's impossible to know if
the discrepancy is a reflection of
problems in the model calculation of the muon capture
or the invalidity of our model assumptions of
fast $\gamma$-decay and pure M1 radiation.
We therefore strongly encourage the 
measurements of $\tau$ and $\delta$ for
the $^{35}$S( $5/2^+$ , 3421)
$\rightarrow$ $^{35}$S( $3/2^+$ , 0) gamma-decay.

\newpage

%
%
\vskip\baselineskip


\noindent $\dagger$ 

Present address: 
Department of Oncology, King Faisal Specialist Hospital and Research Centre, 
Jeddah, KSA, 21499. \\

\noindent $\star$ Present address: The Boeing Company,
Denver Engineering Center, 14261 E. 4th Ave.,
MC AG-00 Bldg 6 Suite 100,
Aurora, CO, USA, 80011 \\

\noindent $\ddagger$ Present address: 
Communications Research Centre, 3701 Carling Ave.  Box 
11490, Station H, Ottawa, Ontario, Canada, K2H 8S2. \\

%
%

%
%

\newpage

\begin{table}
\caption{Gamma--ray yields per $\mu$Cl atom formed
for the cleanly identified lines from $^{35}$Cl($\mu$,$\nu$)$^{35}$S.
Columns 1--5 give the relevant
energies, lifetimes and branching ratios 
from Ref. [20]. The coincident fraction $f_c$ is discussed 
in Sec.\ IIID.}
\label{t-gammas}
\begin{tabular}{lllllll}
 & & & & & & \\
 $E_{\gamma}$ & ( $E$, $J^{\pi}$ )$_i$ & 
($E$, $J^{\pi}$ )$_f$ & B.R. & lifetime & $\gamma$--ray yield & coinc. frac. \\
 (keV)        & (keV)        & (keV)        &
 (\%) &  &
 (per $\mu$Cl atom)  & $f_c$ \\
 & & & & & & \\
\hline
 & & & & & & \\ 
  1991.3 & $(1991, 7/2^-)$ & $(0, 3/2^+)$ &
100 & $(1.02 \pm 0.05) $ns &   ( 7.1$\pm$1.2 ) $\times$10$^{-3}$ & 1.27 $\pm$ 0.06 \\ 
  2347.8 & $(2348, 3/2^-)$ & $(0, 3/2^+)$ &
75 & $0.89 \pm 0.12 $fs &    ( 2.8$\pm$0.5 ) $\times$10$^{-3}$ & 0.77 $\pm$ 0.15 \\ 
  2717.1 & $(2717, 5/2^+)$ & $(0, 3/2^+)$ &
94 & $(70 \pm 25) $fs &    ( 2.9$\pm$0.5 ) $\times$10$^{-3}$ & 1.00 $\pm$ 0.39 \\ 
  2939.6 & $(2939, 3/2^+)$ & $(0, 3/2^+)$ &
100 &  &   ( 5.4$\pm$0.9 ) $\times$10$^{-3}$ & $<$0.30 \\ 
  3421.0 & $(3421, 5/2^+)$ & $(0, 3/2^+)$ &
100 & $<70 fs$ &    ( 5.3$\pm$0.9 ) $\times$10$^{-3}$ & 0.10$\pm$0.15 \\ 
 & & & & & & \\
\end{tabular}
\end{table}

\begin{table}
\caption{The measured capture rates and hyperfine dependences 
for the $^{35}$Cl$(3/2^+ , 0)$ 
$\rightarrow$ $^{35}$S$( 3/2^+ , 2939)$ transition
and the $^{35}$Cl$(3/2^+ , 0)$ 
$\rightarrow$ $^{35}$S$( 5/2^+ , 3421)$ transition.
Note that $\Lambda = 0.12 \Lambda_+ + 0.88 \Lambda_-$
(see Sec. \ref{s-hyperfine} for details).}
\label{t-results}
\begin{tabular}{lll}
 & & \\
  ( $E$, $J^{\pi}$ )$_i$ & $\Lambda$ & 
  $\Lambda_+$/$\Lambda_-$ \\
 (keV)        & ($\times 10^3$ s$^{-1}$)        \
    &         \\
 & & \\
\hline
 & & \\ 
$(2939, 3/2^+)$ &
$12.2 \pm 2.2$  &$0.74 \pm 0.17$ \\ 
$(3421, 5/2^+)$ &
$11.9 \pm 2.2$ & $1.60 \pm 0.19$ \\ 
 & & \\
\end{tabular}
\end{table}


\begin{table}
\caption{Results of our studies of the systematic uncertainties 
in the extraction of the hyperfine dependence
for the $^{35}$Cl($3/2^+$ , 0) $\rightarrow$ $^{35}$S($3/2^+$ , 2939) 
transition.
The input parameters in the time spectrum fits were the 
$t = 0$ position ($t_o$), 
instrumental width ($\sigma$), 
muon disappearance rate ($\Lambda_D$)
and hyperfine transition rate ($\Lambda_h$).
See text for details.
}
\label{t-2939time}
\begin{tabular}{llllll}
                &          &         &       &                  &               \\
 $\Lambda_h$ & $\Lambda_D$ & t$_o$ & $\sigma$ & k & $\Lambda_+ / \Lambda_-$ \\
 $\times 10^6$~s$^{-1}$ & $\times 10^6$~s$^{-1}$ & chans. & chans. & & \\
                &          &         &       &                  &               \\
\hline
                &          &         &       &                  &               \\
    8.0   &    0.225 & 1504.04 & 1.923 &   -0.15$\pm$0.10 & 0.76$\pm$0.16 \\
   10.0   &    0.225 & 1504.04 & 1.923 &   -0.15$\pm$0.10 & 0.76$\pm$0.16 \\
    8.0   &    0.235 & 1504.04 & 1.923 &   -0.18$\pm$0.10 & 0.71$\pm$0.16 \\
   10.0   &    0.235 & 1504.04 & 1.923 &   -0.19$\pm$0.10 & 0.70$\pm$0.16 \\
   10.0   &    0.225 & 1503.84 & 1.923 &   -0.14$\pm$0.10 & 0.78$\pm$0.16 \\
   10.0   &    0.225 & 1504.24 & 1.923 &   -0.16$\pm$0.10 & 0.74$\pm$0.16 \\
   10.0   &    0.225 & 1504.04 & 1.82  &   -0.15$\pm$0.10 & 0.76$\pm$0.16 \\
   10.0   &    0.225 & 1504.04 & 2.02  &   -0.15$\pm$0.10 & 0.76$\pm$0.16 \\
   10.0   &    0.225 & 1504.04 &no tail&   -0.15$\pm$0.10 & 0.76$\pm$0.16 \\
          &          &         &       &                  &               \\
\end{tabular}
\end{table}

\begin{table}
\caption{Results of our studies of the systematic uncertainties 
in the extraction of the hyperfine dependence
for the $^{35}$Cl($3/2^+$ , 0) $\rightarrow$ $^{35}$S($5/2^+$ , 3421) 
transition.
The input parameters in the time spectrum fits were the 
$t = 0$ position ($t_o$), 
instrumental width ($\sigma$), 
muon disappearance rate ($\Lambda_D$)
and hyperfine transition rate ($\Lambda_h$).
See text for details.}
\label{t-3421time}
\begin{tabular}{llllll}
                &          &         &       &                  &               \\
 $\Lambda_h$ & $\Lambda_D$ & t$_o$ & $\sigma$ & k & $\Lambda_+ / \Lambda_-$ \\
 $\times 10^6$~s$^{-1}$ & $\times 10^6$~s$^{-1}$ & chans. & chans. & & \\

                &          &         &       &                  &               \\
\hline
                &          &         &       &                  &               \\
    8.0   &    0.225 & 1504.04 & 1.923 &    0.41$\pm$0.11 & 1.66$\pm$0.18 \\
   10.0   &    0.225 & 1504.04 & 1.923 &    0.40$\pm$0.10 & 1.64$\pm$0.16 \\
    8.0   &    0.235 & 1504.04 & 1.923 &    0.35$\pm$0.10 & 1.56$\pm$0.18 \\
   10.0   &    0.235 & 1504.04 & 1.923 &    0.34$\pm$0.10 & 1.54$\pm$0.18 \\
   10.0   &    0.225 & 1503.84 & 1.923 &    0.34$\pm$0.10 & 1.54$\pm$0.18 \\
   10.0   &    0.225 & 1504.24 & 1.923 &    0.37$\pm$0.10 & 1.59$\pm$0.18 \\
   10.0   &    0.225 & 1504.04 & 1.82  &    0.39$\pm$0.10 & 1.62$\pm$0.18 \\
   10.0   &    0.225 & 1504.04 & 2.02  &    0.39$\pm$0.10 & 1.62$\pm$0.18 \\
   10.0   &    0.225 & 1504.04 &no tail&    0.38$\pm$0.07 & 1.61$\pm$0.11 \\
                &          &         &       &                  &  \\
\end{tabular}
\end{table}

\begin{table}
\caption{
The systematic uncertainties in extracting 
the $\gamma$--$\nu$ correlation coefficient
from the Doppler lineshape 
of the $^{35}$Cl($3/2^+$ , 0) $\rightarrow$ $^{35}$S($3/2^+$ , 2939)
transition.
Note we used a Gaussian instrumental lineshape 
with a centroid $E_o$, width $\sigma$,
and small low-energy tail.
Different polynomials of different powers
were used to explore the sensitivity
to the parameterization of the background
({\it i.e.} $\sum_i x^i$).
See text for details.
}
\label{t-dop2939}
\begin{tabular}{lll}
             &            &                 \\
parameter    &  amount    & correlation $\alpha$ \\
varied       &  changed   &                      \\
             &            &                 \\
\hline
             &            &                 \\
$\sum_i x^i$ &     i = 0.0    &  -0.44$\pm$0.04 \\
$\sum_i x^i$ &     i = 1.0    &  -0.44$\pm$0.04 \\
$\sum_i x^i$ &     i = 2.0    &  -0.44$\pm$0.04 \\
$\sigma$     &   +20.0\%  &  -0.31$\pm$0.04 \\
$\sigma$     &   +10.0\%  &  -0.37$\pm$0.04 \\
$\sigma$     &   -10.0\%  &  -0.49$\pm$0.03 \\
$\sigma$     &   -20.0\%  &  -0.55$\pm$0.03 \\
E$_o$        &     +0.4 ch.  &  -0.33$\pm$0.04 \\
E$_o$        &     +0.2 ch.  &  -0.36$\pm$0.04 \\
E$_o$        &     -0.2 ch.  &  -0.49$\pm$0.04 \\
E$_o$        &     -0.4 ch.  &  -0.54$\pm$0.04 \\
tail         &    +50.0\% &  -0.39$\pm$0.03 \\
tail         &    -50.0\% &  -0.42$\pm$0.03 \\
             &            &                 \\
\end{tabular}
\end{table}

\begin{table}
\caption{
The systematic uncertainties in extracting 
the $\gamma$--$\nu$ correlation coefficient
from the Doppler lineshape 
of the $^{35}$Cl($3/2^+$ , 0) $\rightarrow$ $^{35}$S($5/2^+$ , 3421)
transition.
Note we used a Gaussian instrumental lineshape 
with a centroid $E_o$, width $\sigma$,
and small low-energy tail.
Different polynomials of different powers
were used to explore the sensitivity
to the parameterization of the background
({\it i.e.} $\sum_i x^i$).
See text for details.
}
\label{t-dop3421}
\begin{tabular}{lll}
             &            &                 \\
parameter    &  amount    & correlation $\alpha$ \\
varied       &  changed   &                 \\
             &            &                 \\
\hline
             &            &                 \\
$\sum_i x^i$ &    i = 0.0    &  -0.39$\pm$0.04 \\
$\sum_i x^i$ &    i = 1.0    &  -0.38$\pm$0.04 \\
$\sum_i x^i$ &    i = 2.0    &  -0.39$\pm$0.04 \\
$\sigma$     &   +20.0\%  &  -0.41$\pm$0.04 \\
$\sigma$     &   +10.0\%  &  -0.40$\pm$0.04 \\
$\sigma$     &   -10.0\%  &  -0.39$\pm$0.03 \\
$\sigma$     &   -20.0\%  &  -0.39$\pm$0.03 \\
E$_o$        &     +0.4 ch.  &  -0.39$\pm$0.04 \\
E$_o$        &     +0.2 ch.  &  -0.39$\pm$0.04 \\
E$_o$        &     -0.2 ch.  &  -0.38$\pm$0.04 \\
E$_o$        &     -0.4 ch.  &  -0.39$\pm$0.04 \\
             &            &                 \\
\hline
\end{tabular}
\end{table}

\begin{table}
\caption{Toy model results for the $\gamma$-$\nu$ correlation coefficient
$\alpha$ in the 
$^{35}$Cl$(3/2^+ , 0)$ $\rightarrow$ $^{35}$S$( 5/2^+ , 3421)$
$\rightarrow$ $^{35}$S$( 3/2^+ , 0)$ sequence.
The results labeled `full' corresponds to the complete model calculation
and the results labeled `approx' corresponds to a model calculation
omitting the second forbidden matrix elements.
The quoted values are for $g_a = -1.26$.}
\label{t-toy}
\begin{tabular}{lll}
\hline
& & \\
  $g_p$ &     $\alpha$  & $\alpha$ \\
        &     `full'    & `approx' \\
& & \\
\hline
& & \\
     0  &       -0.240  & -0.216 \\
& & \\
    -4  &       -0.236  & -0.210 \\
& & \\
    -8  &       -0.232  & -0.203 \\
& & \\
    -12 &       -0.228  & -0.196 \\
& & \\
\hline
\end{tabular}
\end{table}

%
%

%

\newpage

\begin{figure}
\begin{center} \leavevmode \epsfysize=0.80\hsize
{\epsfbox{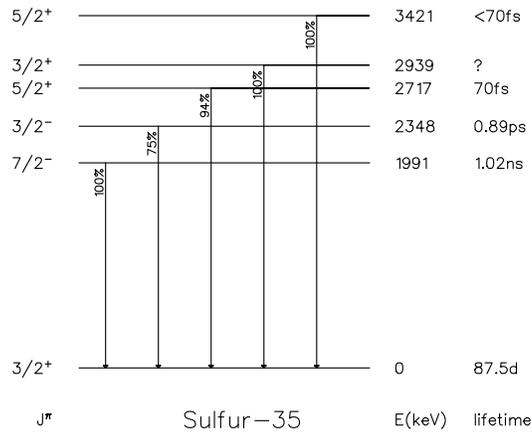}}
\vspace{1.0cm}
\caption{$^{35}$S energy level diagram showing 
cleanly identified $\gamma$--rays.
The energies, spin-parities, lifetimes and branching
ratios were taken from Ref. [20].}
\label{f-gammas}
\end{center}
\end{figure}

\newpage

\begin{figure}
\begin{center} \leavevmode \epsfysize=0.80\hsize
{\epsfbox{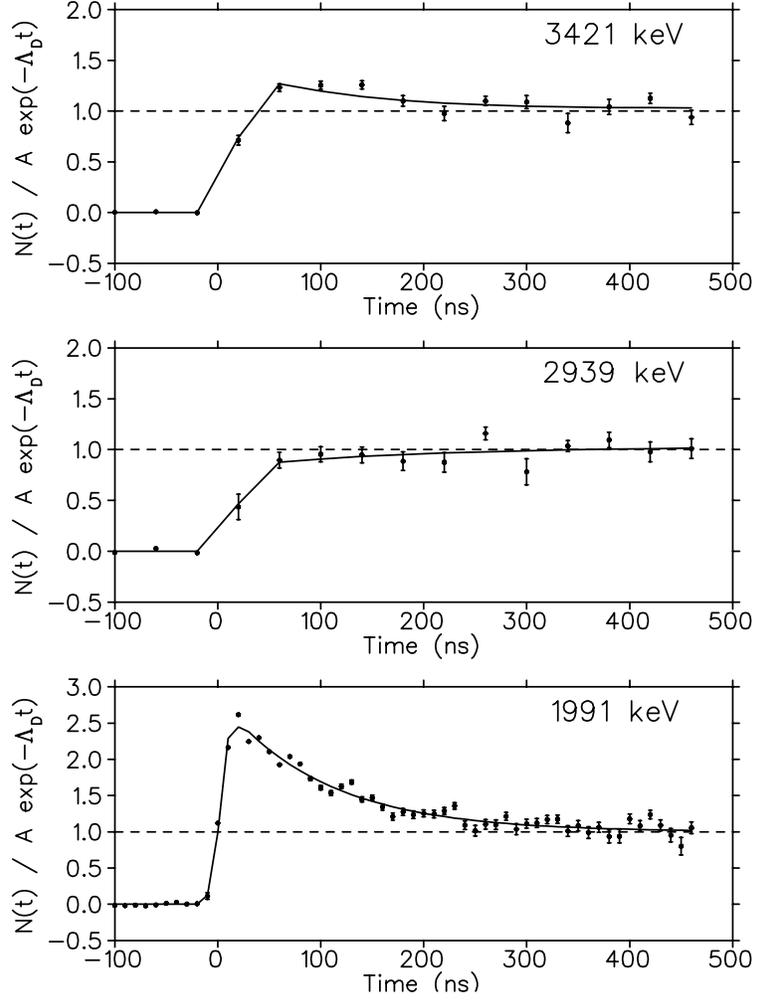}}
\vspace{1.0cm}
\caption{
Time spectra for the 
$1991$ $\gamma$--ray (bottom),
$2939$ $\gamma$--ray (center),
and $3421$ $\gamma$--ray (top).
The points are the experimental data and the solid lines
are the least squares fits.
The muon disappearance rate was ``divided out'' to
better demonstrate the hyperfine effect.
Note the very large hyperfine effect for the
1991~keV $\gamma$--ray was used to fix the
hyperfine transition rate $\Lambda_h$ (see Sec. \ref{s-hyperfine} 
for details).}
\label{f-timefits}
\end{center}
\end{figure}

%

%

%

\newpage

\begin{figure}
\begin{center} \leavevmode \epsfysize=0.80\hsize
 {\epsfbox{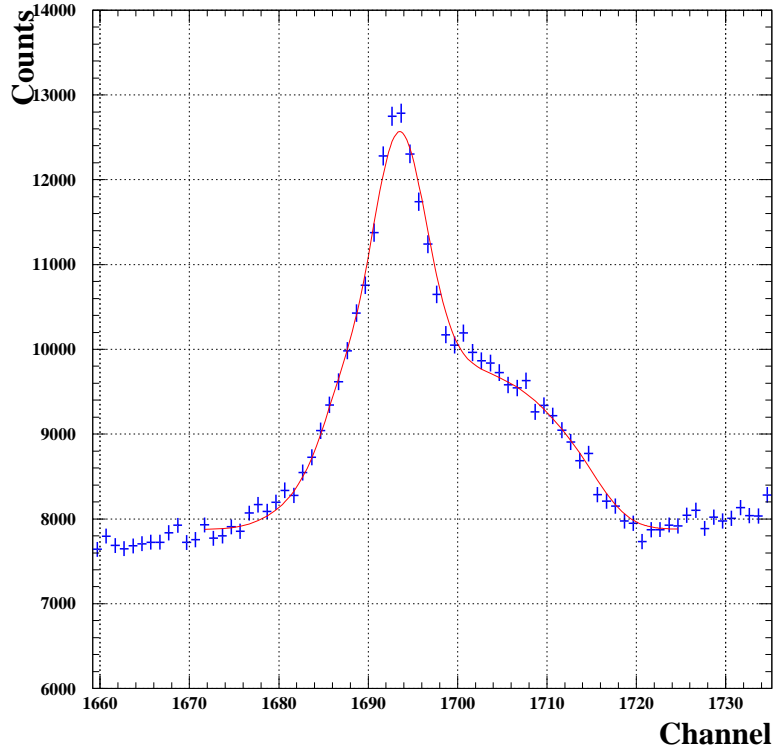}}
\vspace{1.0cm}
\caption{Sample fit to the Doppler broadened energy spectrum 
of the 2939~keV $\gamma$--ray
(the narrower Gaussian-shaped peak 
is a $^{33}$S background line at 2935~keV).}
\label{f-dop2939}
\end{center}
\end{figure}

\newpage

\begin{figure}
\begin{center} \leavevmode \epsfysize=0.80\hsize
 {\epsfbox{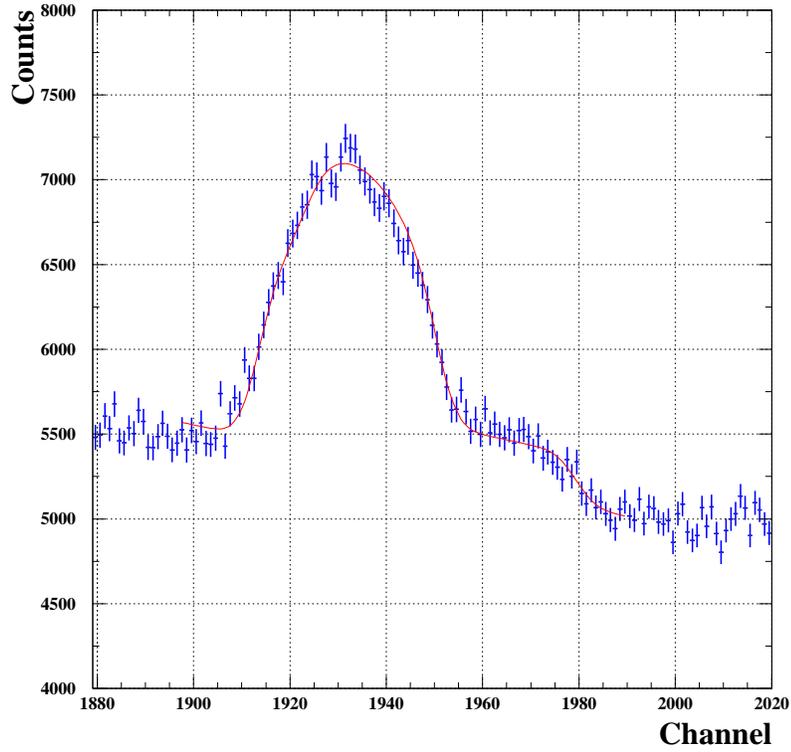}}
\vspace{1.0cm}
\caption{Sample fit to the Doppler broadened energy spectrum 
of the 3421~keV $\gamma$--ray
(the weaker Doppler broadened peak 
is a $^{23}$Ne background line at 3432~keV).}
\label{f-dop3421}
\end{center}
\end{figure}

\newpage

\begin{figure}
\begin{center} \leavevmode \epsfysize=0.80\hsize
\rotate[l] {\epsfbox{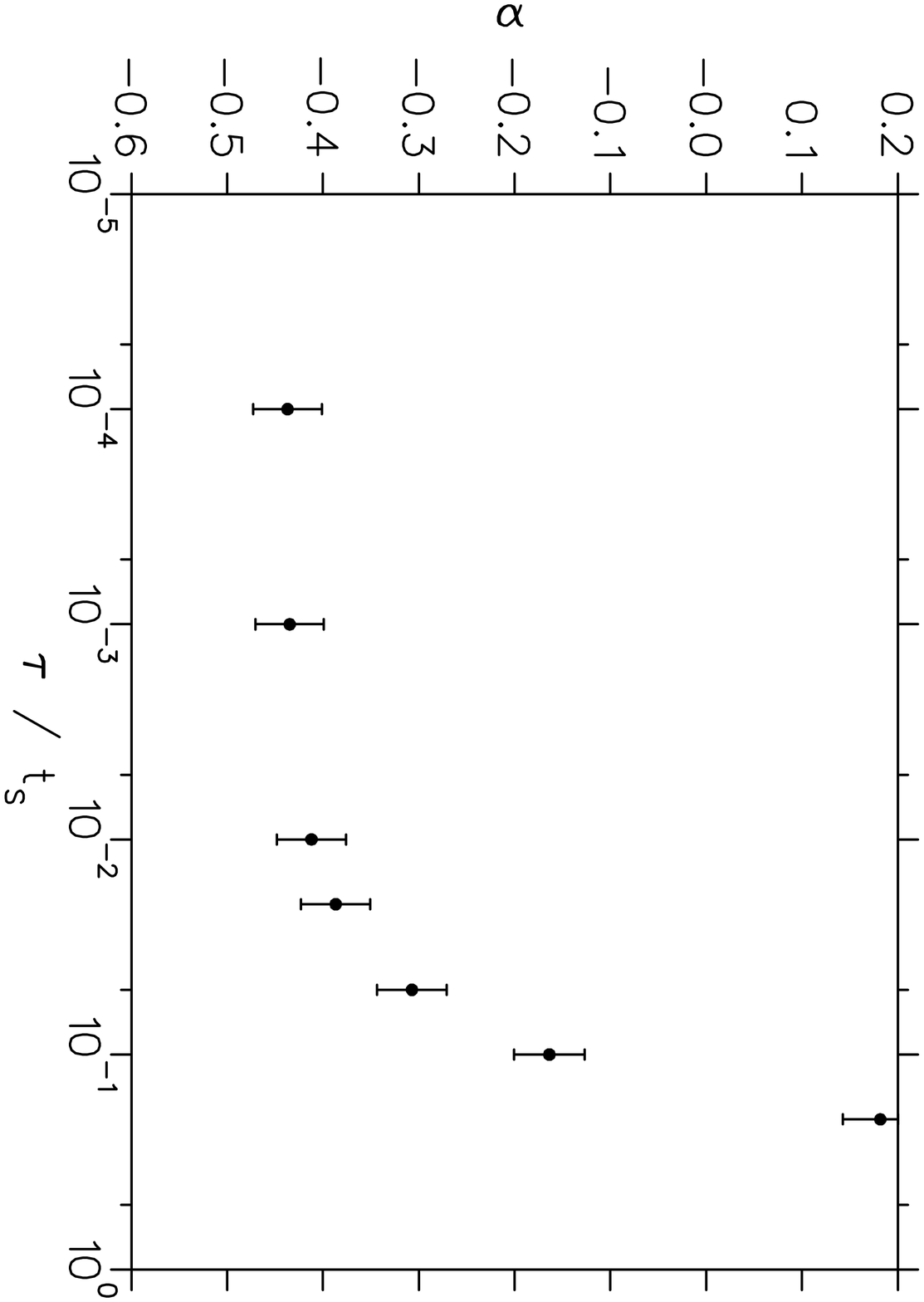}}
\vspace{1.0cm}
\caption{The $\gamma$-$\nu$ angular correlation coefficient $\alpha$
versus the ratio $\tau / t_s$ 
for the $^{35}$S 2939~keV $\gamma$--ray.
Only the statistical errors in $\alpha$ are plotted.}
\label{f-alpha2939}
\end{center}
\end{figure}

\newpage

\begin{figure}
\begin{center} \leavevmode \epsfysize=0.80\hsize
\rotate[l] {\epsfbox{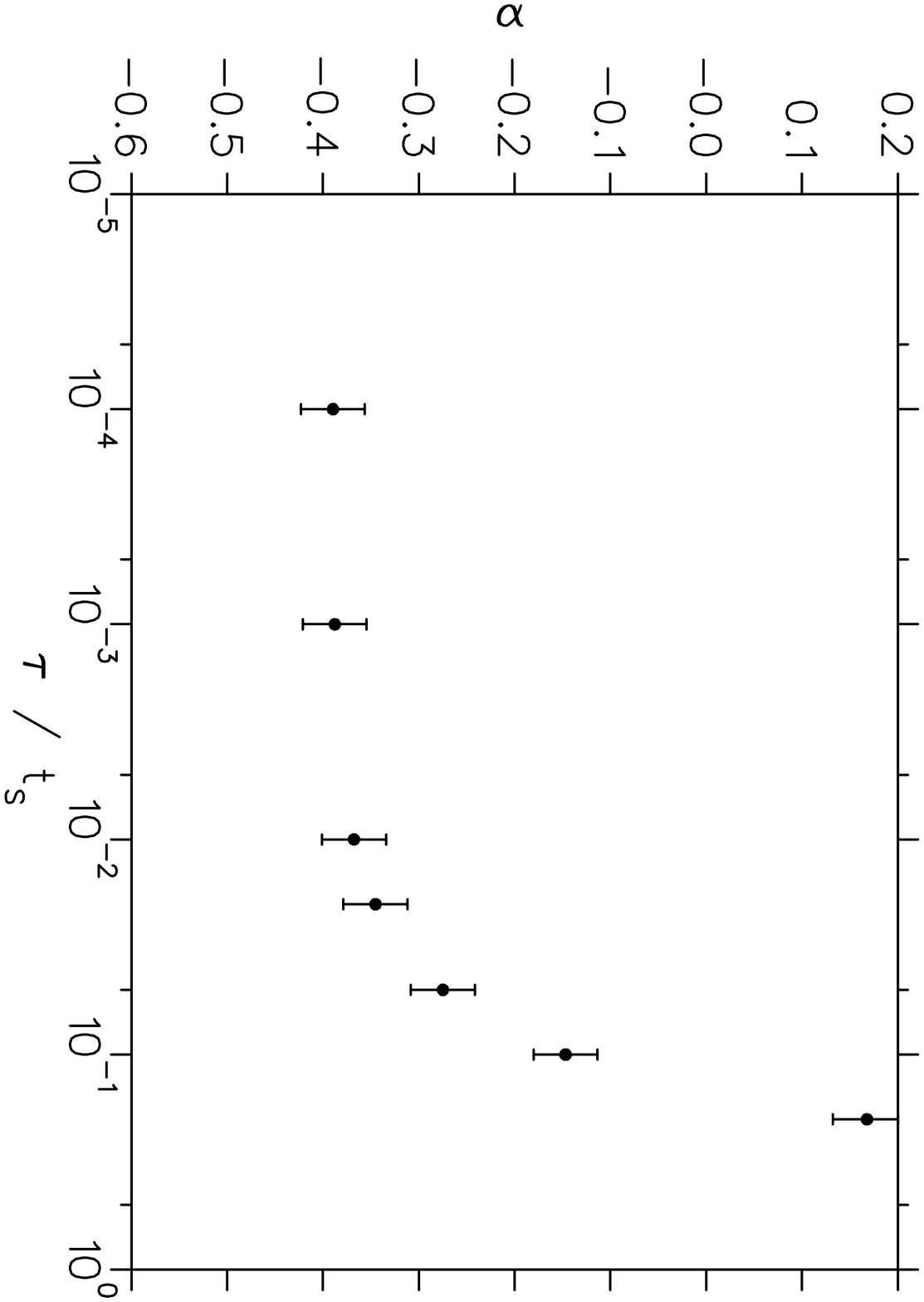}}
\vspace{1.0cm}
\caption{The $\gamma$-$\nu$ angular correlation coefficient $\alpha$
versus the ratio $\tau / t_s$ 
for the $^{35}$S 3421~keV $\gamma$--ray.
Only the statistical errors in $\alpha$ are plotted.}
\label{f-alpha3421}
\end{center}
\end{figure}

\newpage

\begin{figure}
\begin{center} \leavevmode \epsfysize=0.80\hsize
\rotate[l] {\epsfbox{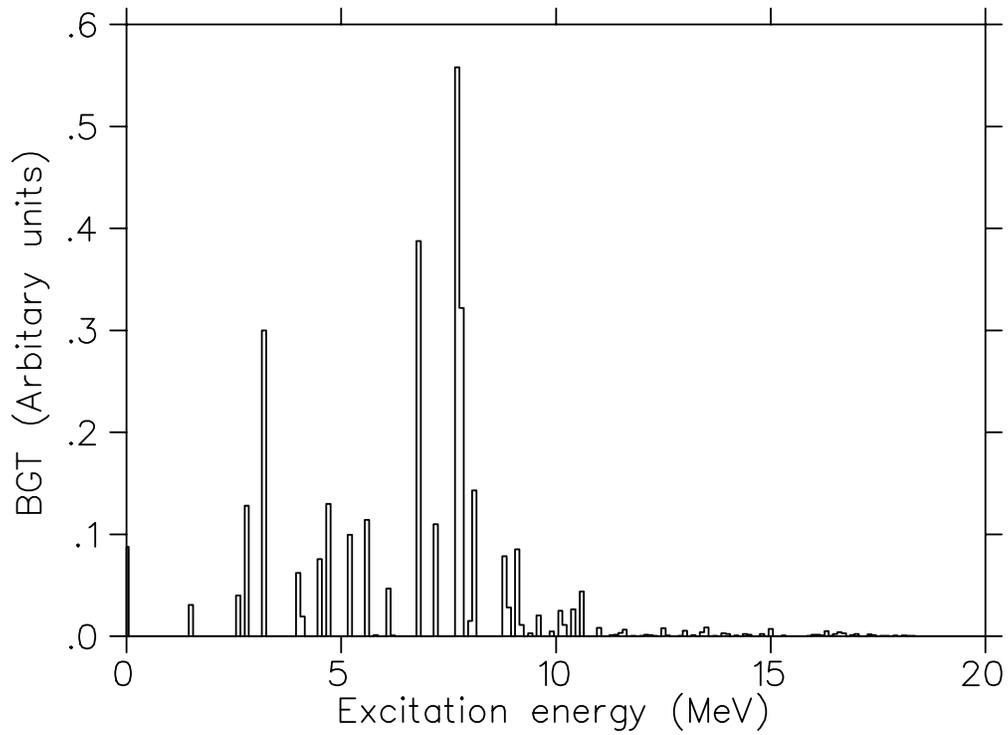}}
\vspace{1.0cm}
\caption{The calculated distribution of 
$^{35}$Cl $\rightarrow$ $^{35}$S allowed GT strength distribution (B$_GT$)
using the 1s--0d shell model and the universal SD interaction
(see text for details).}
\label{f-bgt}
\end{center}
\end{figure}

\newpage

\begin{figure}
\begin{center} \leavevmode \epsfysize=0.80\hsize
\rotate[l] {\epsfbox{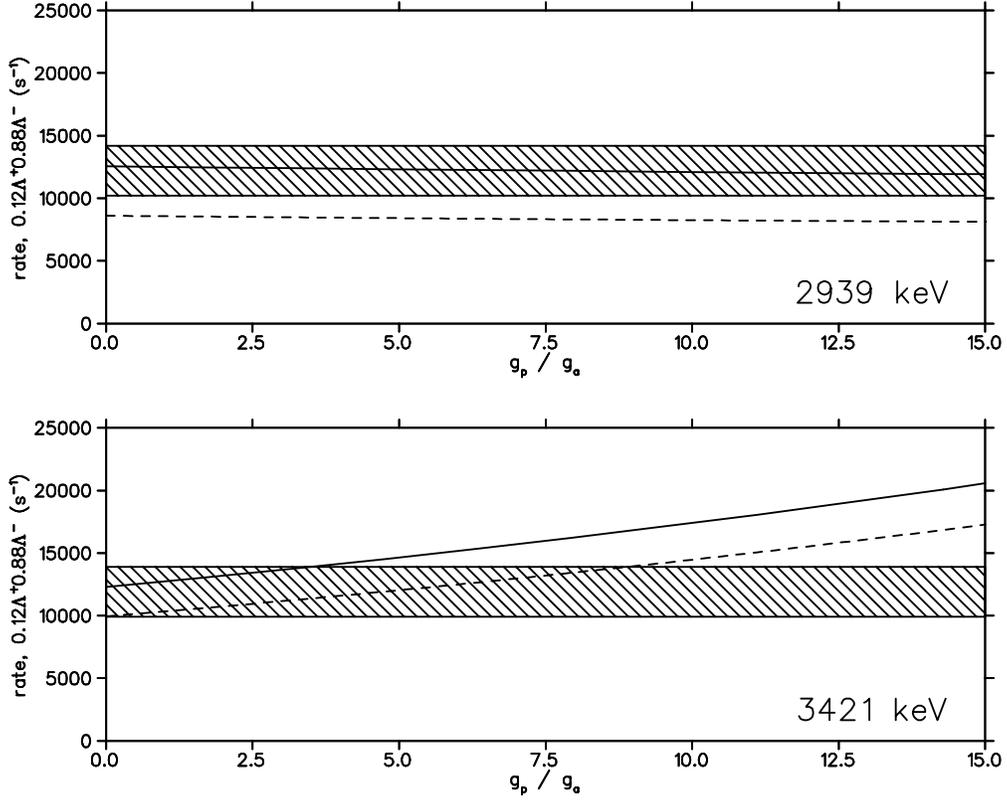}}
\vspace{1.0cm}
\caption{
The calculated capture rate
$\Lambda$ = $0.12 \Lambda_+ + 0.88 \Lambda_-$
for the $^{35}$Cl$(3/2^+ , 0)$ $\rightarrow$ $^{35}$S$( 3/2^+ , 2939)$ 
transition (top)
and the $^{35}$Cl$(3/2^+ , 0)$ $\rightarrow$ $^{35}$S$( 5/2^+ , 3421)$ 
transition (bottom).
The solid lines are for $g_a = -1.26$ and
the dashed lines are for $g_a = -1.00$.
The shaded bands correspond to our experimental results.}
\label{f-rate}
\end{center}
\end{figure}

\newpage

\begin{figure}
\begin{center} \leavevmode \epsfysize=0.80\hsize
\rotate[l] {\epsfbox{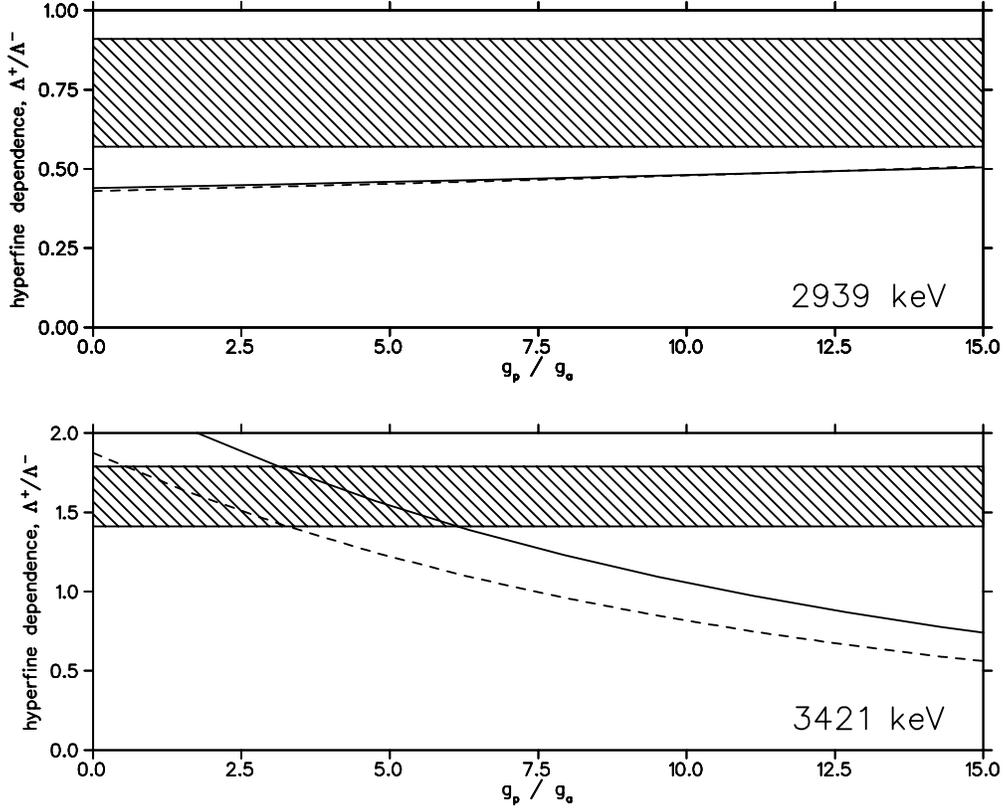}}
\vspace{1.0cm}
\caption{The hyperfine dependence
$\Lambda_+ / \Lambda_-$
for the $^{35}$Cl$(3/2^+ , 0)$ $\rightarrow$ $^{35}$S$( 3/2^+ , 2939)$ 
transition (top)
and the $^{35}$Cl$(3/2^+ , 0)$ $\rightarrow$ $^{35}$S$( 5/2^+ , 3421)$ 
transition (bottom).
The solid lines are for $g_a = -1.26$ and
the dashed lines are for $g_a = -1.00$.
The shaded bands correspond to our experimental results.}
\label{f-hyp}
\end{center}
\end{figure}

\end{document}